\documentclass[showpacs,twocolumn,10pt]{revtex4}
\usepackage{psfig}
\usepackage{graphicx}
\usepackage{dcolumn}
\usepackage{bm}
\usepackage{amsmath}
\usepackage{amsfonts}
\usepackage{amssymb}

\begin{document}

\preprint{}
\title{A Time-Like Naked Singularity}
\author{Rituparno Goswami$^{* a}$, Pankaj S. Joshi$^{* b}$, Cenalo Vaz$%
^{\dag c}$ and Louis Witten$^{\dag d}$}
\affiliation{$^*$Tata Institute for Fundamental Research\\
Homi Bhabha Road, Mumbai 400 005, India\\
$^a$\textrm{Email address: \href{mailto:goswami@mailhost.tifr.res.in}{%
goswami@tifr.res.in}}\\
$^b$\textrm{Email address: \href{mailto:psj@tifr.res.in}{psj@tifr.res.in}}\\
}
\affiliation{$^\dag$ Department of Physics\\
University of Cincinnati, Cincinnati, Ohio, USA.\\
$^c$\textrm{Email address: \href{mailto:vaz@physics.uc.edu}{%
vaz@physics.uc.edu}}\\
$^d$\textrm{Email address: \href{witten@physics.uc.edu}{witten@physics.uc.edu%
}}\\
}

\begin{abstract}
We construct a class of spherically symmetric collapse models in which a
naked singularity may develop as the end state of collapse. The matter
distribution considered has negative radial and tangential pressures, but
the weak energy condition is obeyed throughout. The singularity forms at the
center of the collapsing cloud and continues to be visible for a finite
time. The duration of visibility depends on the nature of energy distribution.
Hence the causal structure of the resulting singularity depends on the nature
of the mass function chosen for the cloud. We present a general model in which
the naked singularity formed is timelike, neither pointlike nor null. Our
work represents a step toward clarifying the necessary conditions for the
validity of the Cosmic Censorship Conjecture.
\end{abstract}

\pacs{04.20.Dw, 04.70.-s, 04.70.Bw}
\maketitle

The cosmic censorship conjecture (CCC) has been widely recognized as one of
the most important open problems in gravitational physics today. This is
because several important areas in the theory and applications of black hole
physics crucially depend on CCC. Nevertheless, the CCC remains unproved and
there exists no mathematically precise and definite statement for the CCC
which one could try to prove (see e.g. [1-6] for some recent reviews, and
references therein).

For this reason a detailed study of dynamically developing gravitational
collapse models within the framework of general relativity becomes rather
essential. The hope is that such a study may allow us to formulate a
provable statement of the CCC, if it is correct in some form. Such
investigations also help us to discard certain statements of the CCC which
might sound plausible but for which there exist counter-examples which show
that the CCC cannot be valid in such a form. They may even illustrate the
physical conditions that give rise to naked singularities (NS) or black
holes (BH) as end states of a realistic gravitational collapse. So far, such
dynamical collapse studies have focused largely on collapse models that
create either BH or NS, depending on the nature of the initial profiles of
density, pressure, and velocity from which the collapse develops. In many of
these cases, when a NS develops, it is located at the center of the
spherically symmetric cloud (a central singularity, see e.g. [5-7]). In that
case, there will exist families of non-spacelike future directed geodesics,
which will be accessible to distant observers in the future, and which will
terminate at the singularity in the past, thus making it visible in
principle. This is opposed to the BH case where the apparent horizon forms
early enough to cover all of the singularity, with no portion of it
remaining visible to outside observers.

If we require the pressure to be positive then the "central" singularity, if
it is naked, corresponds to a singularity along a visible null line. The
remainder of the singularity is spacelike and covered by a horizon. In this
paper, however, we permit the pressure to be negative and examine the
structure of the singularity. We construct an explicit solution in which the
singularity may be timelike. It may even change its character, being
timelike along a certain region and, after being visible for a finite time,
turning spacelike and being covered. The collapsing matter is described by a
particularly chosen matter field that satisfies the weak energy condition
although the radial and tangential pressures are negative and unequal. While
what is presented here is a specific construction of a class of collapse
models, involving somewhat special choices, we make sure that physical
reasonability conditions such as the energy conditions and the regularity of
the initial data at the initial surface are respected.

The spherically symmetric metric in a general form can be written as
\begin{equation}
ds^{2}=-e^{2\nu(t,r)}dt^{2} + e^{2\psi(t,r) }dr^{2} + R^{2}(t,r)d\Omega^{2},
\label{eq:metric}
\end{equation}
where $d\Omega^{2}$ is the line element on a two-sphere. Choosing the
comoving frame, the stress-energy tensor for a general (type I) matter field
is given in a diagonal form as
\begin{equation}
T^{t}_{t}=-\rho;\; T^{r}_{r}= p_{r};\; T^{\theta}_{\theta}=T^{\phi}_{\phi}=
p_{\theta},  \label{eq:setensor}
\end{equation}
where $\rho$, $p_{r}$ and $p_{\theta}$ are the energy density, and the
radial and tangential pressures respectively. We assume that the matter
field satisfies the \textit{weak energy condition}, that is, the energy
density as measured by any local observer is non-negative and, for any
timelike vector $V^{i}$,
\begin{equation}
T_{ik}V^{i}V^{k}\ge0
\end{equation}
which amounts to,
\begin{equation}
\rho\ge0;\; \rho+p_{r}\ge0;\; \rho+p_{\theta}\ge 0.
\end{equation}
The initial data consists of three metric functions, the energy density, and
the radial and tangential pressures at the initial time $t=t_{i}$. This is
given in terms of six arbitrary functions of the radial coordinate, \textit{%
viz.,} $\nu(t_{i},r)=\nu_{0}(r)$, $\psi (t_{i},r)=\psi_{0}(r)$, $R(t_{i},r)=r
$, $\rho(t_{i},r)=\rho_{0}(r)$, $p_{r}(t_{i},r)=p_{r0}(r)$, $p_{\theta}
(t_{i},r)=p_{\theta0}(r)$, where, using the scaling freedom for the radial
co-ordinate $r$ we have chosen $R(t_{i},r)=r$ at the initial epoch. The
dynamic evolution of the initial data is then determined by the Einstein
equations, which for the metric (\ref{eq:metric}) become ($8\pi G=c=1$),
\begin{align}
\rho=\frac{F^{\prime}}{R^{2}R^{\prime}}, &~~ p_{r}=-\frac{\dot{F}}{R^{2}\dot{%
R}}  \label{eq:ein1}
\end{align}
\begin{equation}
\nu^{\prime}=\frac{2(p_{\theta} -p_{r})}{\rho+p_{r}}\frac{R^{\prime}}{R}-
\frac{p_{r}^{\prime}} {\rho+p_{r}}  \label{eq:ein2}
\end{equation}
\begin{equation}
-2\dot{R}^{\prime}+R^{\prime}\frac{\dot{G}}{G}+\dot{R}\frac{H^{\prime}}{H}=0
\label{eq:ein3}
\end{equation}
\begin{equation}
G-H=1-\frac{F}{R},  \label{eq:ein4}
\end{equation}
where $F=F(t,r)$ is an arbitrary function. In spherically symmetric
spacetimes $F(t,r)$ is interpreted as the mass function, with $F\ge 0$. In
order to preserve the regularity of the initial data we must also require $%
F(t_{i},0)=0$, \textit{i.e.,} the mass function should vanish at the center
of the cloud. The functions $G$ and $H$ are defined as $G(t,r)=e^{-2%
\psi}(R^{\prime})^{2}$ and $H(t,r)=e^{-2\nu } (\dot{R})^{2}$.

All the initial data above are not mutually independent: from equation (\ref%
{eq:ein2}) we find that the function $\nu_{0}(r)$ is determined in terms of
rest of the initial data. Also, by rescaling of the radial coordinate $r$,
the number of independent initial data functions reduces to four. We then
have a total of five field equations with seven unknowns, $\rho$, $p_{r}$, $%
p_{\theta}$, $\psi$, $\nu$, $R$, and $F$, giving us the freedom of choice of
two free functions. Selection of these functions, subject to the given
initial data and weak energy condition, determines the matter distribution
and metric of the space-time and thus leads to a particular collapse
evolution of the initial data. At this point it is convenient to introduce a
scaling variable $v(t,r)$, defined as
\begin{equation}
R(t,r)=rv(t,r),  \label{eq:R}
\end{equation}
where,
\begin{equation}
v(t_{i},r)=1,~~ v(t_{s}(r),r)=0~~ \mathrm{and}~~ \dot{v}<0,
\end{equation}
the last condition being necessary for a collapse. Let us consider the
following choice of the allowed free functions, $F(t,r)$ and $\nu(t,r)$,
\begin{equation}
F(t,r)=r^{3}\mathcal{M}(r)v,  \label{eq:mass}
\end{equation}
where $r^3\mathcal{M}(r)$ is a suitably differentiable and monotonically
non-decreasing function, and
\begin{equation}
\nu(t,r)=\nu_{0}(R).  \label{eq:nu}
\end{equation}
The function $\mathcal{M}$ may be expanded in a Taylor series about $r=0$,
\begin{equation}
\mathcal{M}(r)=\mathcal{M}_{0}+\mathcal{M}_{2}r^{2}+\mathcal{M}%
_{3}r^{3}+\cdots.  \label{eq:M}
\end{equation}
Then, from equation (\ref{eq:ein1}), we have
\begin{equation}
\rho=\frac{3\mathcal{M}v+r[\mathcal{M}_{,r}v+\mathcal{M}v^{\prime}]}{%
v^{2}[v+rv^{\prime}]}  \label{eq:rho1}
\end{equation}
and
\begin{equation}
p_{r}=-\frac{\mathcal{M}(r)}{v^{2}}.  \label{eq:pr1}
\end{equation}
The above choice of mass function therefore imples that the radial pressure
is negative. The \textit{weak energy condition}, however, does hold. If $%
R^{\prime}=v+rv^{\prime}$ and $F^{\prime}$ are both positive in the equation
(\ref{eq:rho1}), then we clearly $\rho\ge 0$. Again, for $\rho + p_r \geq 0$
at all epochs, it must be true that
\begin{equation}
(2\mathcal{M}_{0}+4\mathcal{M}_{2}r^{2}+5\mathcal{M}_{3}r^{3}+\cdots)v\geq 0.
\label{eq:condition}
\end{equation}
But, because $v\geq 0$, it follows that if the condition $\rho+p_r \geq 0$
is satisfied at the initial epoch, it is satisfied throughout the evolution.
Finally, from equation (\ref{eq:ein2}),
\begin{equation}
\rho+p_{\theta}=\frac{1}{2}(\rho+p_r)\left[1+R\nu_0(R)_{,R}\right] +\frac{r^2%
\mathcal{M} }{R^2}\ge0  \label{eq:condition2}
\end{equation}
if $\left[1+R\nu_0(R)_{,R}\right]\ge0$ for all epochs. This provides a
necessary condition for the weak energy condition to be satisfied.

At the initial epoch we then have
\begin{equation}
\rho_{0}(r)=3\mathcal{M}_{0}+5\mathcal{M}_{2}r^{2}+6\mathcal{M}%
_{3}r^{3}+\cdots  \label{eq:rho2}
\end{equation}
and
\begin{equation}
p_{r_{0}}(r)=-[\mathcal{M}_{0}+\mathcal{M}_{2}r^{2}+\mathcal{M}%
_{3}r^{3}+\cdots]  \label{eq:pr2}
\end{equation}
At the initial epoch, the radial and the tangential pressures must be equal
at the center and all the pressure gradients must vanish. It follows that
the initial tangential pressure must have the form
\begin{equation}
p_{\theta_{0}}(r)=-[\mathcal{M}_{0}+p_{\theta_{2}}r^{2}+p_{\theta_{3}}r^{3}+%
\cdots].  \label{eq:pt1}
\end{equation}
Hence from equation (\ref{eq:ein2}) we see that $\nu_{0}(r)$ becomes,
\begin{equation}
\nu_{0}(r)=a_{2}r^{2}+a_{3}r^{3}+\cdots  \label{eq:nu0}
\end{equation}
where
\begin{equation*}
a_{2}=\frac{p_{\theta_{2}}+\mathcal{M}_{2}}{2\mathcal{M}_{0}},~~ a_{3}=\frac{%
p_{\theta_{3}}+\mathcal{M}_{3}}{2\mathcal{M}_{0}}
\end{equation*}
and, from equation (\ref{eq:nu}),
\begin{equation}
\nu(t,r)=\nu_{0}(R)=a_{2}R^{2}+a_{3}R^{3}+\cdots  \label{eq:nu1}
\end{equation}
The dynamic evolution of $p_{\theta}(t,r)$ is obtained by inserting equation
(\ref{eq:nu}) in equation(\ref{eq:ein2}) and simplifying to get,
\begin{equation}
p_{\theta}(r,v)=p_{r}+\frac{Rp_{r}^{\prime}}{2R^{\prime}} +\frac{1}{2}%
\nu_{0}(R)_{,R}R(\rho+p_{r})  \label{eq:ptrt}
\end{equation}
There therefore exists an $\epsilon$ ball around the central shell for which
$p_{\theta}=p_{r}$ and the perfect fluid equation of state is valid.

Using equation (\ref{eq:nu}) in equation (\ref{eq:ein3}), we get,
\begin{equation}
G(t,r)=b(r)e^{2\nu_{0}(R)},  \label{eq:G}
\end{equation}
where $b(r)$ is another arbitrary function of $r$. In corresponding dust
models, we can write $b(r)=1+r^{2}b_{0}(r)$, where $b_{0}(r)$ is the energy
distribution function of the collapsing shells. Thus, the metric (\ref%
{eq:metric}) becomes,
\begin{equation}
ds^{2}=-e^{2(a_{2}R^{2}+\cdots)}dt^{2}+\frac{R^{\prime2}e^{-2(a_{2}R^{2}+%
\cdots)}dr^{2}}{1+r^{2}b_{0}(r)}+R^{2}d\Omega^{2}  \label{eq:metric1}
\end{equation}
and is valid for small values of $r$, for all epochs, \textit{i.e.} for all
values of $v(r,t)$, till the singularity.

Solving the equation of motion (\ref{eq:ein4}) we find that
\begin{equation}
\dot{v}=-e^{2\nu_{0}(rv)}\sqrt{v^{2}(2a_{2}+2a_{3}rv\cdots)+b_{0}(r)e^{2%
\nu_{0}(rv)}+\mathcal{M}(r)},  \label{eq:vdot}
\end{equation}
which may be integrated to obtain
\begin{equation}
t(v,r)=\int_{v}^{1}{\frac{e^{-2\nu_{0}(rv)}dv}{\sqrt{v^{2}(2a_{2}+2a_{3}rv%
\cdots)+b_{0}(r)e^{2\nu_{0}(rv)}+\mathcal{M}(r)}}}.  \label{eq:sing1}
\end{equation}
We note that the radial coordinate $r$ is treated as a constant in the above
equation, which gives the time taken for a shell labeled $r$, to reach a
later epoch $v$ in collapse from the initial epoch $v=1$.

It is clear that an explicit solution of the above integral will give a
closed form solution of the form $t=f(v,r)$ or, inversely, $v=g(t,r)$, which
will then determine the metric function $R=rg(t,r)$ thereby giving an exact
solution for the metric (\ref{eq:metric1}). Unfortunately, the integral
cannot be expressed in closed form and we make a Taylor expansion of the
integral about the center of the cloud.
\begin{equation}
t(v,r)=t(v,0)+r\mathcal{X}(v)+O(r^{2})  \label{eq:scurve41}
\end{equation}
where the function $\mathcal{X} (v)$ is given by,
\begin{equation}
\mathcal{X}(v)=-\frac{1}{2}\int_{v}^{1}dv\frac{2v^{4}a_{3}+b_{1}}{\left[
\sqrt{b_{0}(0)+2v^{2}a_{2}+\mathcal{M}(0)}\right] ^{3}}  \label{eq:tangent11}
\end{equation}
If a closed form solution of $R$ exists up to the first approximation, it
will be of the form $R=r\mathcal{X}^{-1}[(t(v,r)-t(v,0))/r]$. Therefore, by
expanding as above we are actually solving for $R$ and so for the metric (%
\ref{eq:metric1}) to the first approximation, although we do not write it in
closed form. This is because it is only the sign of $\mathcal{X} (0)$ that
determines the final end state of the collapse, which is the issue of
interest here.

The time taken for the central shell at $r=0$ to reach the singularity, $t_s(0)$,
is given by
\begin{equation}
t_{s}(0)=\int_{0}^{1}\frac{dv}{\sqrt{v^{2}(2a_{2})+b_{0}(0)+\mathcal{M} (0)}}%
\label{eq:sing2}
\end{equation}
The time taken for the other shells ($r\neq 0$) to reach the singularity,
$t_s(r)$, can be given as,
\begin{equation}
t_{s}(r)=t_{s}(0)+r\mathcal{X} (0)+O(r^{2}),  \label{eq:scurve4}
\end{equation}
where the function $\mathcal{X} (0)$ is given by,
\begin{equation}
\mathcal{X} (0)=-\frac{1}{2}\int_{0}^{1}dv\frac{2v^{4}a_{3} +b_{1}} {\left[
\sqrt{b_{0}(0)+2v^{2}a_{2}+\mathcal{M} (0)}\right] ^{3}}  \label{eq:tangent1}
\end{equation}
We see that by suitable choice of the coefficients of initial density,
pressure and energy profiles we can make $\mathcal{X} (0)$ positive or
negative. Furthermore, at the singularity, for a constant $v$ surface we
have
\begin{equation}
\lim_{v\rightarrow0}v^{\prime}=\sqrt{b_{0}(r)+\mathcal{M} (r)}\mathcal{X}
(0)+\mathcal{O}(r^{2})  \label{eq:vdash}
\end{equation}
and, because we have expressions for $v^{\prime}$ and $\dot{v}$, near the
central singularity, we can in principle calculate $v(r,t)$ in the
neighborhood of the central singularity. This solves the system of Einstein
equations.

The apparent horizon is the boundary of the trapped region of the space-time
and is given by $R/F=1$. If the neighborhood of the center gets trapped
earlier than the singularity, then it will be covered and a black hole will
be the final state of the collapse. Otherwise, the singularity can be naked
with non-spacelike future directed trajectories escaping from it to outside
observers. Using (\ref{eq:mass}), we find that the apparent horizon is just
the surface
\begin{equation}
r^{2}\mathcal{M}(r)=1.
\end{equation}
Therefore, if $r_{b}^{2}\mathcal{M}(r_{b})<1$ there will be no trapped
surfaces in spacetime, where $r_{b}$ is the radial coordinate of the
boundary of the cloud.

It is simplest to examine the nature of the singularity by noting that it
occurs at $R=0$. This implies that at the singularity,
\begin{equation}
ds^{2}=[\exp(2\psi)-\exp(2\nu)\frac{R^{{\prime}^{2}}}{\dot{R}^{2}}]dr^{2}.
\end{equation}
If the right hand side is negative the singularity is timelike. Therefore,
for a timelike singularity, $G-H>0$, or
\begin{equation}
1-r^{2}\mathcal{M}(r)>0.
\end{equation}
But, because the function $r^{2}\mathcal{M} (r)$ is monotonically
non-decreasing, it follows that the singularity is timelike near $r=0$,
becomes null at $r^{2}\mathcal{M} (r)=1$ and finally spacelike when $r^{2}%
\mathcal{M} (r)>1$.

This result may also be obtained explicitly near the center by examining the
outgoing null geodesics. To see this specifically at $r=0$, the outgoing
radial null lines are given by
\begin{equation}
\frac{dt}{dr}=e^{\psi -\nu },  \label{eq:null1}
\end{equation}%
which, at the singularity, corresponds to
\begin{equation}
\left( \frac{dt}{dr}\right) _{null}=\lim_{v\rightarrow 0}\frac{rv^{\prime }}{%
\sqrt{1+r^{2}b_{0}(r)}}.  \label{eq:null2}
\end{equation}%
In order to find the existence or otherwise of an outgoing null geodesic
from the singularity we substitute the value of $v^{\prime }$ at the
singularity in the above equation to obtain
\begin{equation}
\left( \frac{dt}{dr}\right) _{null}=\left[ \frac{r\sqrt{b_{0}(r)+\mathcal{M}%
(r)}}{\sqrt{1+r^{2}b_{0}(r)}}\right] \left( \frac{dt}{dr}\right) _{s}.
\label{eq:null3}
\end{equation}%
If
\begin{equation}
\left( \frac{dt}{dr}\right) _{s}=\mathcal{X}(0)\geq 0
\end{equation}%
then, for all values of $r$ for which
\begin{equation}
\left[ \frac{r\sqrt{b_{0}(r)+\mathcal{M}(r)}}{\sqrt{1+r^{2}b_{0}(r)}}\right]
\leq 1
\end{equation}%
or $1-r^{2}\mathcal{M}(r)>0$ , there will be a visible outgoing null
geodesic leaving the singularity.

The model discussed here is based on a special choice of the mass function.
The fluids described by this choice are not intended to describe an actual
physical system: both the radial and the tangential pressures are negative.
Nevertheless the initial data satisfy appropriate regularity conditions and
the weak energy condition is maintained throughout. Therefore, the system
can serve as a guide to further clarifying and precisely formulating the CCC.

Understanding what is possible in dynamically developing collapse models is
necessary to arrive at a plausible concrete statement of the CCC. We have an
example that shows that even in spherically symmetric collapse, the naked
singularity need not always be either pointlike or null, but can have an
interesting causal structure, including being timelike.

\end{document}